# KOSMOS and COSMOS: New facility instruments for the NOAO 4-meter telescopes

Paul Martini*[a], J. Elias[b], S. Points[c], D. Sprayberry[b], M.A. Derwent[a], R. Gonzalez[a], J.A. Mason[a], T.P. O'Brien[a], D.P. Pappalardo[a], R.W. Pogge[a], R. Stoll[a,d], R. Zhelem[a,e], P. Daly[b], M. Fitzpatrick[b], J.R. George[b], M. Hunten[b,f], R. Marshall[b], G. Poczulp[b], S. Rath[b], R. Seaman[b], M. Trueblood[b,g], K. Zelaya[b]

[a]Department of Astronomy, The Ohio State University, 140 West 18th Avenue, Columbus, OH 43221 USA; [b]National Optical Astronomy Observatory, 950 North Cherry Street, Tucson AZ 85719 USA; [c]Cerro Tololo Inter-American Observatory, Casilla 603, La Serena, Chile; [d]Current Address: Department of Astronomy, Yale University, New Haven, CT 06511 USA; [e]Current Address: Australian Astronomical Observatory; PO Box 915; North Ryde NSW 1670 Australia; [f]Current Address: University of Arizona, Lunar and Planetary Laboratory, Tucson, AZ 85721 USA [g]Current Address: Winer Observatory, P.O. Box 797, Sonoita, AZ 85637 USA

## ABSTRACT

We describe the design, construction and measured performance of the Kitt Peak Ohio State Multi-Object Spectrograph (KOSMOS) for the 4-m Mayall telescope and the Cerro Tololo Ohio State Multi-Object Spectrograph (COSMOS) for the 4-m Blanco telescope. These nearly identical imaging spectrographs are modified versions of the OSMOS instrument; they provide a pair of new, high-efficiency instruments to the NOAO user community. KOSMOS and COSMOS may be used for imaging, long-slit, and multi-slit spectroscopy over a 100 square arcminute field of view with a pixel scale of 0.29 arcseconds. Each contains two VPH grisms that provide R~2500 with a one arcsecond slit and their wavelengths of peak diffraction efficiency are approximately 510nm and 750nm. Both may also be used with either a thin, blue-optimized CCD from e2v or a thick, fully depleted, red-optimized CCD from LBNL. These instruments were developed in response to the ReSTAR process. KOSMOS was commissioned in 2013B and COSMOS was commissioned in 2014A.

**Keywords:** Spectrographs, Imagers, Astronomical Instruments, Multi-Object Spectroscopy, VPH, NOAO, Mayall, Blanco

## 1. INSTRUMENT OVERVIEW

The Ohio State University and the National Optical Astronomy Observatories (NOAO) developed the KOSMOS and COSMOS instruments to be new facility instruments for the 4-m Mayall telescope at Kitt Peak National Observatory and the 4-m Blanco telescope at the Cerro Tololo Inter-American Observatory. These instruments were built to provide the user community with modern, high-efficiency spectrographs that meet many of the scientific needs described in the ReSTAR (Renewing Small Telescopes for Astronomical Research) Report[1]. KOSMOS was commissioned at the Mayall telescope in October 2013 and COSMOS was commissioned at the Blanco telescope in April 2014.

In order to expedite the development, as well as to minimize total costs, the KOSMOS and COSMOS instruments are largely based on OSMOS[2,3] (Ohio State Multi-Object Spectrograph), which was commissioned at the 2.4m Hiltner telescope of the MDM Observatory in April 2010. The heritage from OSMOS includes an all-refractive optical design that enables imaging, long-slit, and multi-slit spectroscopy over a wide field, the capability for rapid, repeatable reconfiguration between observing modes, and sufficient capacity to have a wide range of slits, filters, and dispersers mounted simultaneously. KOSMOS and COSMOS were designed to be self-contained instruments that mount to the telescope in a single module that only requires power, an internet connection, and a fiber connection for the detector system.

## 2. OPTICS

KOSMOS and COSMOS were designed to mount at the Cassegrain focus of the Mayall and Blanco telescopes. The focal ratio at the Cassegrain focus is approximately f/7.9 and the scale is 6.6 arcseconds per mm with very mild curvature (R=3.33m). The optical design for these telescopes are nearly identical, and as a consequence KOSMOS and COSMOS have the same optical design. The instruments reimage an f/7.9 input beam almost 12 arcminutes (103mm) in diameter onto a 2048x4096 CCD with 15μm pixels at a plate scale of 0.29 arcseconds per pixel. The wavelength range is from 360nm to 1000nm.

### 2.1 Optical Design and Image Quality

The KOSMOS and COSMOS optical layout is shown in Figure 1, and the optical prescription is listed in Table 1. The collimator for KOSMOS and COSMOS is a 5-element, f/7.9 Double Gauss design with a 14 degree field of view. This collimator is identical to the one in OSMOS, and the complete collimator lens barrel assembly was procured from the same vendor, Jenoptik (formerly Coastal) Optical Systems in Jupiter, FL. The collimator doublet was bonded with Sylgard 184, a silicon elastomer. The collimated beam diameter is 54mm, and the pupil is 68mm from the vertex of the last collimator element.

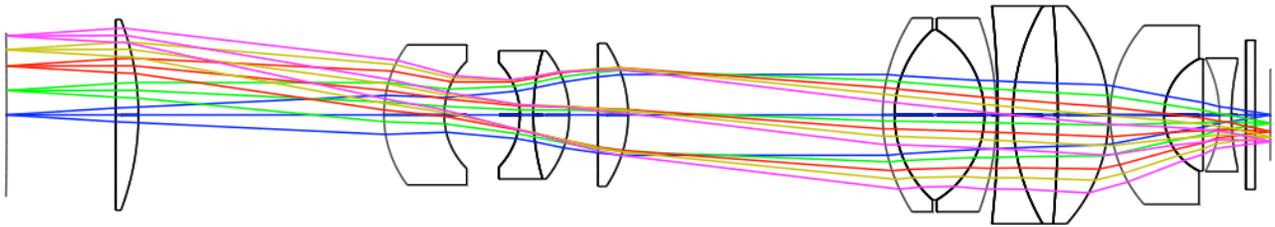

Figure 1. Optical layout for KOSMOS and COSMOS. The telescope focal surface is on the far left and the detector is on the far right. The slit wheel is coincident with the telescope focal surface and may contain long slits and multi-slit masks for spectroscopy. Dispersers and filters are located in the 170mm wide collimated space between lenses 5 and 6.

The camera is substantially faster than the one in OSMOS. The camera is a 9-element, f/2.7 Petzval design with a 26 degree field of view. The first surface of the camera is aspheric. The camera contains two triplets and one doublet. We planned to procure the complete camera barrel assembly from the same optics vendor, but the bonded elements partially debonded when thermally tested at the low temperatures required for winter, nighttime operations on Kitt Peak. We consequently took delivery of the individual optical elements and developed a fluid-coupled solution for the two triplets and doublet. This development is described elsewhere in these proceedings[4].

Due to the large number of air-glass interfaces, particular care was paid to the quality of the coatings. Evaporated Coatings, Inc. in Willow Grove, PA coated the lenses. Witness samples indicate that the coatings average less than 1% reflectivity from 350nm to 1000nm.

### 2.2 Spectroscopy

The camera was designed to have a wide field of view in order provide broad wavelength coverage in spectroscopic mode. Figure 2 illustrates how the almost 12 arcminute diameter field of view is cropped to 100 square arcminutes by the 2048x4096 detector system, such that the maximum length of the long slits is 10 arcminutes. The current long slits include widths of 0.6, 0.9, 1.2, and 1.5 arcseconds. All may be placed at the center of the field or offset by ±160 arcseconds to adjust the wavelength range on the detector and of peak diffraction efficiency of the Volume Phase Holographic (VPH) gratings employed in the dispersers. Multi-object spectroscopy is feasible over the entire field of view accessible to imaging observations.

Each instrument was commissioned with an identical pair of dispersers**.** Each dispersing element is a VPH grating bonded between a pair of prisms (a VPH grism) that maintain the zero-deviation optical path in spectroscopic mode. The VPH gratings are 90mm square and bonded between a pair of NBK7 flats such that the total thickness of each grating is 10mm. The fringe orientation is nominally parallel to the height dimension. One VPH grism is optimized for bluer light (hereafter the Blue VPH) and the other is optimized for redder light (hereafter the Red VPH).

The Blue VPH has a fringe frequency of 1172 lines per mm and is bonded between a pair of PBM2Y prisms with apex angles of 26.44 degrees. The resolution with a 1 arcsecond wide slit varies from about 2000 to 2600 from the blue to the red end. The prisms are coated with an anti-reflection coating that is optimized for 350 to 700nm.

The Red VPH has a fringe frequency of 842 lines per mm and is bonded between a pair of PBM2Y prisms with apex angles of 28.6 degrees. The resolution with a 1 arcsecond wide slit varies from about 2000 to 2600 from the blue to the red end. The prisms are coated with an anti-reflection coating that is optimized for 500nm to 1000nm.

Table 2. Optical prescription for KOSMOS and COSMOS. The collimator doublet is bonded with a thin layer of Sylgard 184. The camera triplets and doublet are coupled with a fluid. The front surface of the first camera triplet (CAM-1) is an even asphere ($r^4 = -7.16678 \times 10^{-8}$, $r^6 = -2.08578 \times 10^{-12}$).

### KOSMOS Optical Prescription

| Element | Lens | Surface | Radius | Thickness | Element | Diameter |
|---|---|---|---|---|---|---|
| Slit Mask | | 0 | 3330 | 73 | | 103 |
| Field Lens | COL-1 | 1 | Infinity | 15 | BSL7Y | 128 |
| | | 2 | -180.89 | 164 | | 128 |
| Meniscus | COL-2 | 3 | 78.8 | 40.3 | BSM51Y | 94 |
| | | 4 | 50.681 | 50.5 | | 72 |
| | COL-3 | 5 | -50.681 | 9 | BAL51Y | 86 |
| Collimator doublet | COL-4 | 6 | 152.3 | 24 | CAF2 | 86 |
| | | 7 | -67.245 | 19.285 | | 86 |
| Collimator singlet | COL-5 | 8 | Infinity | 20 | CAF2 | 86 |
| | | 9 | -86.26 | 170 | | 86 |
| Camera Triplet #1 | CAM-1 | 10 | 111.278 | 8 | BSM51Y | 114 |
| | CAM-2 | 11 | 76.919 | 60 | CAF2 | 114 |
| | CAM-3 | 12 | 67.019 | 8 | BAL35Y | 114 |
| | | 13 | 187.006 | 1 | | 130 |
| Camera Triplet #2 | CAM-4 | 14 | 662.224 | 10 | BAL35Y | 146 |
| | CAM-5 | 15 | 143.802 | 30 | CAF2 | 146 |
| | CAM-6 | 16 | 869.959 | 34 | BSM51Y | 146 |
| | | 17 | 109.182 | 1 | | 146 |
| Camera Doublet | CAM-7 | 18 | 91 | 36 | BSM51Y | 120 |
| | CAM-8 | 19 | 41.990 | 26 | CAF2 | 75.2 |
| | | 20 | Infinity | 11 | | |
| Field Flattener | CAM-9 | 21 | 80.721 | 8 | PBM18Y | 85 |
| | | 22 | 154.755 | 9.568 | | 70 |
| Dewar window | CAM-10 | 23 | Infinity | 6.35 | SILICA | 100 |
| | | 24 | Infinity | 10 | | 100 |
| CCD | | 25 | Infinity | 0 | | 61 x 31 |

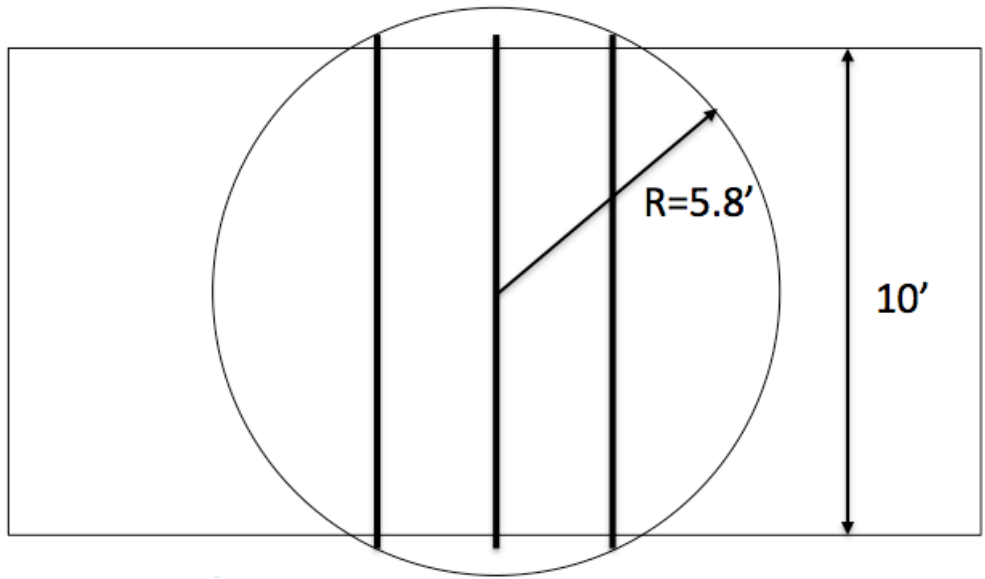

Figure 2. Schematic diagram of the imaging field of view (circle) and the detector geometry (rectangle). The optics correct an 11.6 arcminute diameter field of view, of which 10 arcminutes illuminates the detector in the spatial direction. This also sets the maximum length of the long slits. Light is dispersed along the long axis of the detector.

## 3. MECHANICAL DESIGN

The mechanical design of KOSMOS and COSMOS is very similar to OSMOS[2,3]. Figure 3 shows a photo of KOSMOS mounted on the 4-m Mayall telescope at the Kitt Peak National Observatory and Figure 4 shows a photo of the interior of COSMOS. All three instruments have the same optical bench layout, aperture wheels, nearly identical collimator and camera focus mechanisms, and enclosure. The main differences are a slight decrease in the overall length that accommodates the faster camera, the shutter, the addition of telescope adapter plates, and modifications to the exit plate to accommodate different detector systems. There are also numerous performance, reliability, assembly, and maintenance improvements.

KOSMOS and COSMOS are identical except in two respects. The first is that the telescope adapter plates have slightly different thicknesses to accommodate the different back focal distances of their respective telescopes. The second is that the exit plate of each instrument contains mounting features appropriate to the Dewar style used at each site.

### 3.1 Instrument Layout

The instrument enclosure was designed to provide structural stiffness for the instrument and contains an integrated optical bench. The Instrument Electronics Box (IEB) is mounted to a cantilevered wing on the entrance plate of the enclosure. This decouples the mass of the IEB from the remainder of the instrument and minimizes flexure. A counterweight is mounted on the opposing wing of the entrance plate to maintain rotational balance about the optical axis of the instrument. The IEB is to the right of the main instrument enclosure in Figure 3, while the counterweight is the scarlet square to the left. The detector system Dewar is attached to the exit plate of the instrument and the detector controller electronics are attached to the Dewar. The entire instrument can be mounted and operated on its handling cart. In addition to moving the instrument around the observatory, the instrument may be pivoted between zenith and horizon pointing for telescope installation and service procedures, respectively.

The nominal light path through the instrument can be followed with Figure 4. The light path begins at the entrance plate, which includes a manual dark hatch that may be closed during the day and when the instrument is off the telescope. The shutter is immediately behind the entrance plate, and thus the dark hatch provides protection for the shutter blades. The next unit is the six-position slit wheel. This wheel normally contains one open position for imaging mode, while the remaining five may contain either long slits or multi-slit masks. The collimator is immediately after the slit wheel. The collimator is mounted in a V-block that is attached to a linear drive focus mechanism. There is a 170mm space between the collimator and camera. This space contains a six-position disperser wheel and two six-position filter wheels. The camera is also mounted in a V-block that is attached to a linear drive focus mechanism. The end of the camera is approximately coincident with the exit plate of the enclosure, and the detector system is mounted to the exit plate.

All of the mechanisms in KOSMOS have been designed to operate over a temperature range of $-10^0$F to $+100^0$F. Most of the aperture wheel components, including motors, sensors, gears, hinges, detent assemblies, and filter cell assemblies are identical to components that have been used on previous instruments built by The Ohio State University Imaging Sciences Laboratory that are located at MDM Observatory, Cerro Tololo, and other observatories around the world. The focus mechanisms are nearly identical to several in the Multi-Object Double Spectrograph[5] at the Large Binocular Telescope.

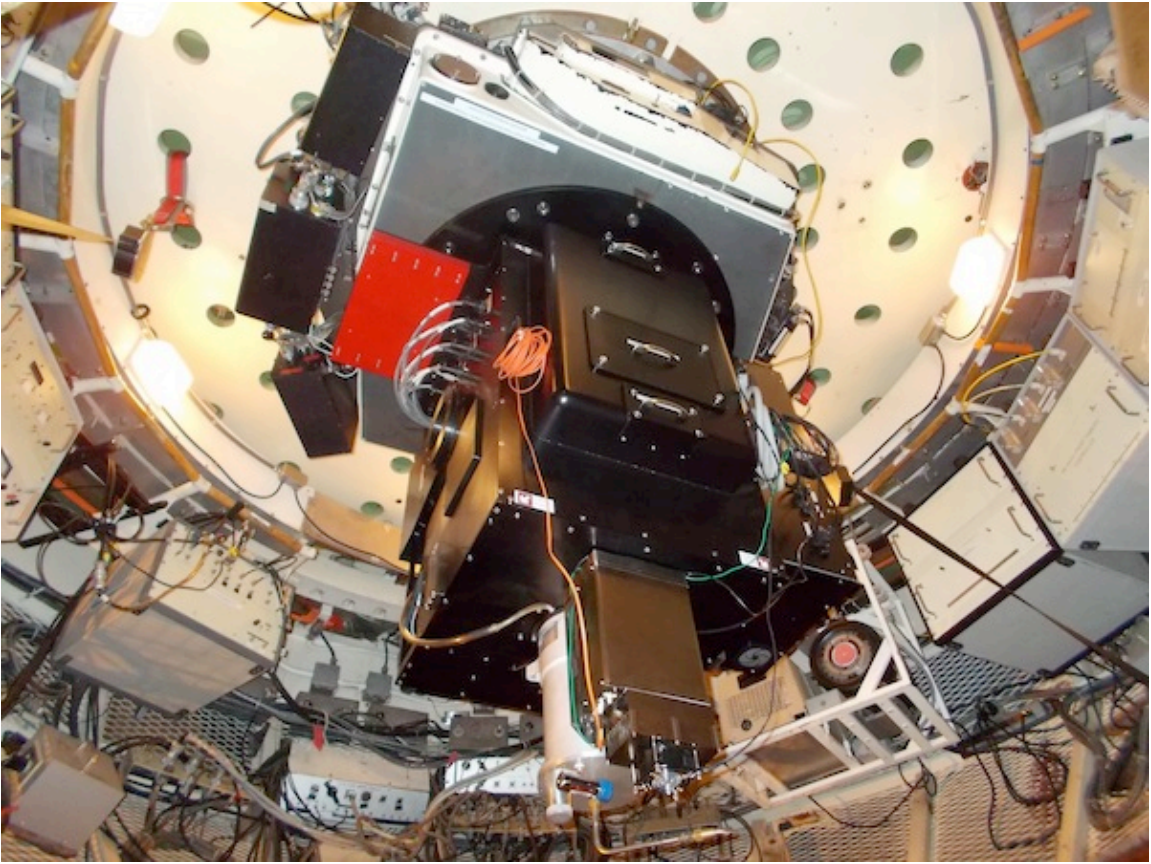

Figure 3. KOSMOS mounted on the 4-m Mayall telescope at Kitt Peak National Observatory. The round, black circle is the telescope adapter. The instrument is mounted to this adapter, and handles for the slit wheel, disperser wheel, and filter wheel access covers are visible from top to bottom along the front of the instrument. This is referred to as the top panel, which describes its orientation when the instrument is pointing at the horizon. The KPNO e2v Dewar is mounted at the bottom of the instrument (to the exit plate), along with the Torrent controller electronics. The instrument electronics box is mounted on the right side of the instrument, while the scarlet square on the left is the counterweight for the instrument electronics.

### 3.2 Aperture Wheels

All four aperture wheels have identical drive components, and are identical to those in OSMOS. They are based on a long heritage of successful aperture wheel designs used in many other instruments. The aperture wheels are direct driven by stepper motors with gear and pinion rotary mechanisms. Inductive proximity sensors are used to determine if each wheel is in position, and the identity of the current position. A spring loaded, roller detent mechanically docks the wheels in position. The holding margin of the detent is nearly a factor of three times more than required. The docking repeatability for the wheels has been measured to be better than 1.5μm from on-sky testing with a pinhole mask mounted in the slit wheel. The maximum time to change between apertures in any of the wheels is less than 10 seconds. All four wheels may be repositioned simultaneously.

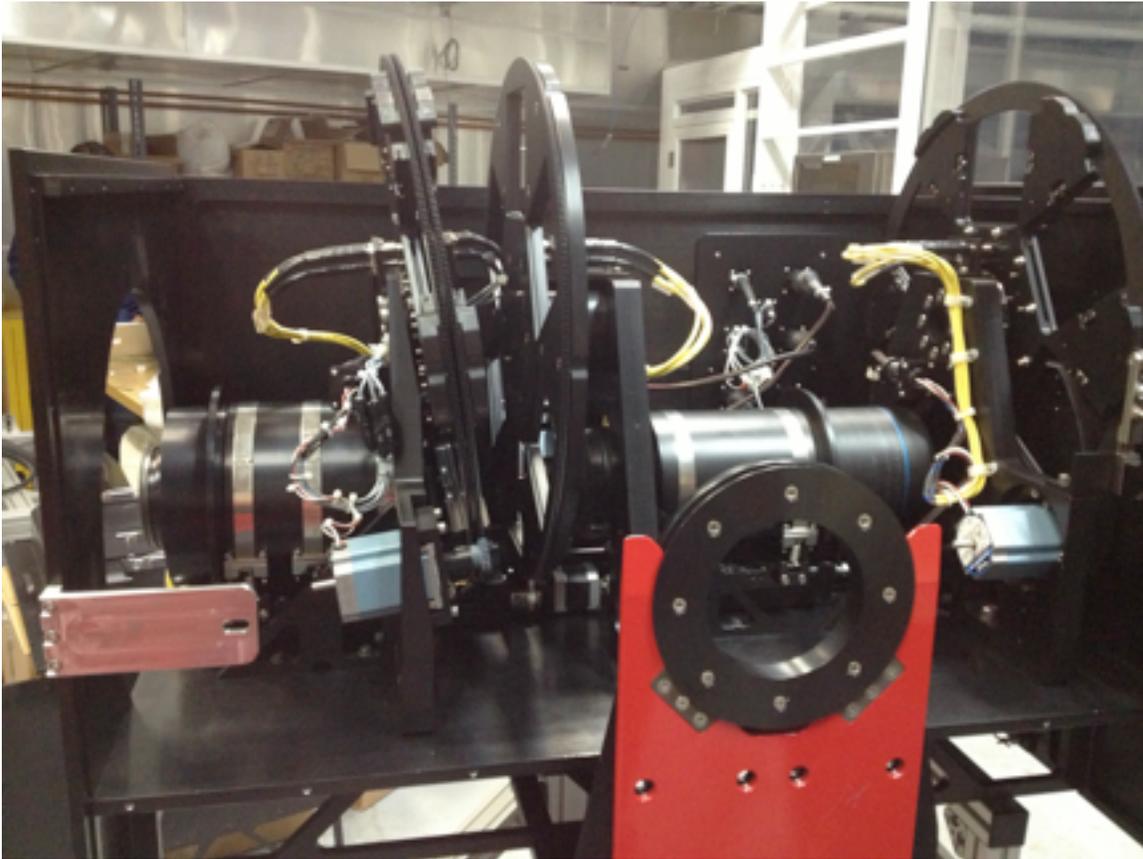

Figure 4. COSMOS on its handling cart. The instrument is shown in the 'horizon pointing' orientation without one side cover, the top cover, and the detector system. The instrument entrance plate is on the right and the detector system mounts at the large, circular aperture on the left. The four aperture wheels from right to left are the slit wheel, disperser wheel, and two (tilted) filter wheels. The shutter (not visible) is between the slit wheel and the entrance plate. The collimator and camera lens barrels are mounted on linear stages.

The only difference between the wheels is their aperture cells. The slit wheel and filter wheels use radially inserted, cartridge style cells, while the disperser wheel uses a face-mounted cell. There are two types of slit wheel cells: multi-slit and long slit. The multi-slit cells clamp a round disc of stainless steel with laser cut slits into the mask holder cell, while the long slits use an interchangeable cell that has been precision cut by a micro-EDM process. The filter cells are almost identical to the multi-slit mask cells, with the exception that they are square and can hold a maximum filter size of 100mm x 100mm x 9.5mm.

The disperser cells are made of titanium in order to match the CTE of the materials used to make the dispersers. The dispersers are epoxied to the cells with Scotch-Weld 2216 gray, two-part epoxy. The disperser cells are mounted to the disperser wheel with eight captive screws. Dowel pins in the wheel prevent the cell from being inserted incorrectly. The disperser wheel contains features to mount counterweights for the dispersers to maintain the balance of the wheel.

The two filter wheels are tilted by eight degrees relative to the other two wheels. This tilt makes it unlikely that parasitic (ghost) reflections off a filter will be reimaged on the detector. This tilt produces a slight wavelength shift to the blue in the bandpass of interference filters, and is significant for narrowband filters.

The aperture wheels can be easily accessed via the three removable panels in the top cover of the instrument enclosure (visible in Figure 3). Only a few minutes are required to exchange slit masks or filters, while exchange of dispersers should require on order ten minutes. These operations do not require removal of the instrument from the telescope. At present, we expect to routinely perform slit changes during the night, but filter and disperser changes will occur only during daytime, in order to minimize risk to filters and dispersers.

## 3.3 Focus Mechanisms

The collimator and camera lens barrels are mounted to identical linear focus mechanisms that are designed for repeatability and stability. The linear focus mechanisms consist of a THK linear actuator that is driven by a stepper motor and 1:100 harmonic gear drive combination. The harmonic reducer increases the cogging torque far beyond what is required to hold the focus stage in place under the gravity load expected at zenith. Inductive proximity sensors are located at each end of the stage and provide homing and end-of-travel limit functions. The sensors are adjusted to provide ±1.5mm of travel from the nominal focus. A V-block attached to the bearing plate of the actuator supports the optics barrel. Spring loaded, steel straps hold the barrel in the V-block with approximately 5G of holding capacity. Spherical washer sets are used between the barrel flange and the end face of the V-block to prevent the bolted connection from biasing the engagement of the barrel in the V-block. This mounting scheme allows the lens barrels to be removed from the instrument without loss of optical alignment.

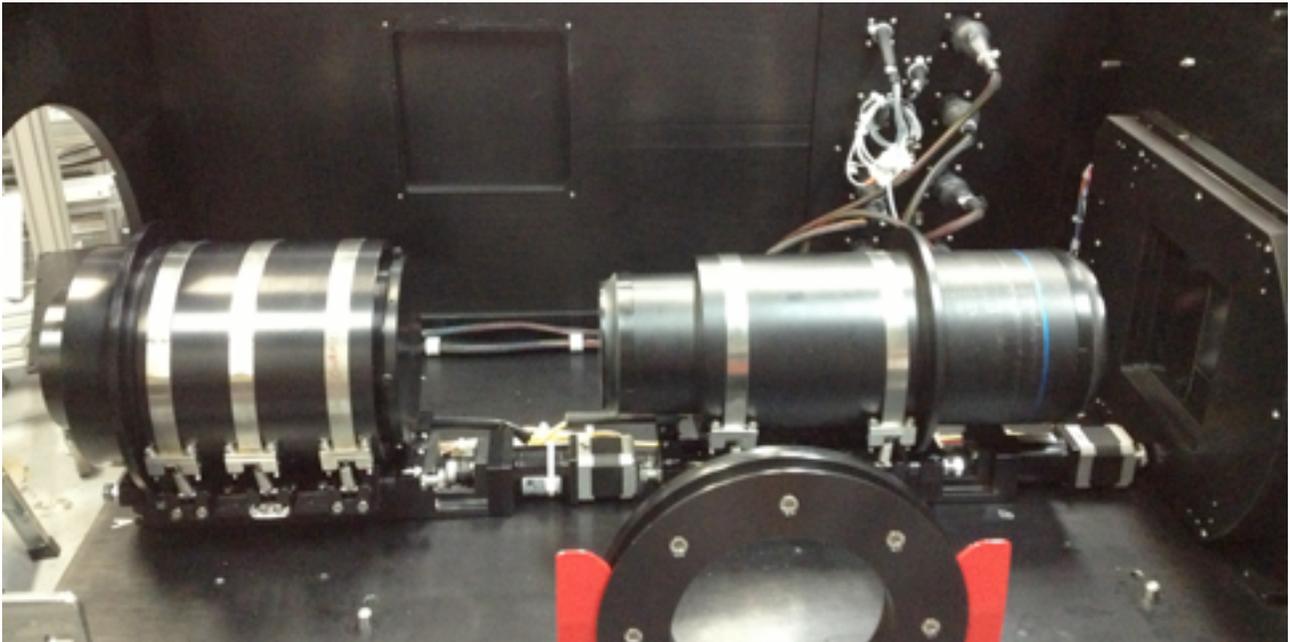

Figure 5. COSMOS with just the camera (left) and collimator (right) lens barrels installed on their respective focus mechanisms on the optical bench. The camera barrel has three spring-loaded straps due to its greater weight. Several precision dowel pins for the wheel mechanisms are visible in the bench. The PS500 shutter is visible to the far right.

Due to the close proximity of the camera barrel to the Dewar window, and the very large drive torque generated by the harmonic drive, we replaced the helical coupling that was used in OSMOS with an R+W torque-limiting coupling. The R+W SK2 series torque-limiting coupling connects the output shaft of the harmonic drive to the ball screw of the THK actuator and has an adjustable 0.1Nm – 0.6Nm torque range. Our couplings came preset at 0.4Nm, which means that once the torque required to turn the ball screw exceeds 0.4Nm, the coupling will disengage and rotate freely until the load is released or the drive direction is reversed. Should the end of travel limit sensor fail, this feature should prevent the linear stage from shearing off the sensor bracket and potentially causing damage to the Dewar window or some other internal component.

One major development goal was to decrease the flexure measured with OSMOS[2], particularly as the faster camera is more massive than the OSMOS camera. Some of the OSMOS flexure was traced to the linear actuator, and so for KOSMOS and COSMOS we acquired a "precision grade" (THK KR3306B+200LPO-1100) model that is rated for nearly three times higher stiffness for the moment about the lead screw axis. We conducted a series of measurements that simulated the instrument under full gravity load with the instrument on its side and horizon pointing to determine the maximum flexure associated with the 32-pound camera barrel acting perpendicular to the bearing guides. Loads were applied to the side of the barrel with a digital force gauge while readings were obtained with a digital indicator mounted on the opposite side of the barrel. This test showed that the full side gravity load deflection was only 35μm, which is half the deflection measured for the standard grade actuator used in OSMOS.

## 3.4 Optical Bench and Enclosure

The optical bench supports the four aperture wheels and two focus mechanisms, and serves as the bottom plate of the enclosure. Ribs run beneath the bench to improve stiffness, as well as serve as cable trays for the mechanism cables that run from the patch panel (shown below the scarlet counterweight in Figure 3) to the IEB. Locating pins in the bench allow any of the mechanisms to be extracted from the instrument for service and later re-installed without the need for realignment.

The optical bench is different from the OSMOS bench in only three respects. First, the optical bench is made from a solid plate of 6061-T6 aluminum with no light-weighting features. In contrast, the OSMOS bench contains many light-weighting pockets on its bottom face due to weight limitations at the Hiltner telescope. Second, even though the mechanism spacing of all three instruments are identical, the KOSMOS and COSMOS benches are two inches shorter due to the faster camera. Finally, the internal cable tie-down zones were relocated for improved serviceability and better mechanism access.

The enclosure is constructed from a series of aluminum plates that provide mounting features, structural stiffness, and protection from dust and stray light. In addition to the optical bench, there are rigid entrance and exit plates that contain mounting features for the telescope adapter, the IEB and its counterweight, and the detector system. There are also left and right side panels, and a top panel. The top and both side panels may be removed to better service the instrument on its cart. The top panel includes three access ports for exchanging slit masks, dispersers, and filters. The top panel may also be removed while the instrument is mounted on the telescope. All panels have notched joints and overlapping seams to help make the instrument light tight.

The light tightness of the instrument enclosure was a significant design consideration, particularly once it became clear that the shutter could only fit at the entrance plate without a significant redesign. The access hatches in the top cover were consequently designed to ensure substantial overlap to the frames of these hatches, as well as the panel for instrument mechanism cables, in order to minimize light leaks.

Since the 4-m telescopes have a much larger payload capacity than the Hiltner telescope, we investigated the impact of removing many of the light-weighting features that are part of the OSMOS optical bench and enclosure. We used the finite element analysis tools in Solidworks Simulation to compare instrument flexure for three scenarios. These were a fully light-weighted optical bench and enclosure panels (like OSMOS), a solid optical bench with light-weighted enclosure panels, and a solid bench and solid enclosure panels. We monitored X, Y, and Z displacements of nodes located at the collimator, camera, and detector while applying gravity loads directly in the X, Y, or Z directions. Once the data were collected for all instrument orientations, we calculated the image shift.

As expected, the analysis predicted an improvement in stiffness with the removal of light-weighting features. The fully light-weighted instrument would have an image shift of nearly one pixel due to flexure of the instrument structure, while removal of the light weighting from the optical bench and enclosure would reduce the image motion to less than ½ pixel. Removal of light weighting from the optical bench alone would result in an instrument weight increase of 10 pounds, while removal of light weighting from both the optical bench and the enclosure would increase the instrument weight by 56 pounds. We ultimate chose to manufacture a solid optical bench and maintain light-weighted enclosure panels. The weight difference between a light-weighted optical bench and a solid optical bench is minimal, yet this produced a substantial increase in stiffness. In contrast, a decrease of only 1/10 of a pixel of image motion was predicted with the addition of solid enclosure panels.

## 3.5 Telescope Interface

The instruments mount to the rotator-guider modules at their respective telescopes, and use the facility guiding and calibration units. The main telescope interface tasks were to build a telescope adapter that would align the instrument to the telescope optical axis and provide the correct back focal distance, confirm that the instruments would fit within the Cassegrain cages, and determine the most convenient way to move the instruments to the Cassegrain cages of their respective telescopes.

The back focal distance of the Mayall and Blanco telescope is approximately twice the 3-inch back focal distance of the Hiltner telescope. The Mayall and Blanco telescopes also differ by approximately one quarter of an inch. Adapters with different thicknesses were constructed for the two telescopes to allow use of the same entrance plate hole pattern and optical bench design as OSMOS. The telescope adapters were constructed from bolted 6061-T6 aluminum parts.

We analyzed the stiffness of the telescope adapter with Solidworks Simulation. For a 1G gravity load, the analysis predicts a 2μm translation of the instrument and a 10μm tip of the instrument, which corresponds to an image shift of 4.5μm. The corresponding pupil shift is predicted to be 20μm. Since the pivot point for the rotation of the instrument is so close to the telescope focal plane, the focus shift is essentially zero. There is also an approximately 3μm displacement along the optical axis with a 1G gravity load when pointed at the zenith. This shift is about an order of magnitude smaller than the camera's depth of focus.

When mounted in the zenith orientation on their carts, KOSMOS and COSMOS have a 49x40 inch footprint and are 69 inches tall. As a result, each fits on the Observatory service elevator. The height on the cart was designed such that the instrument could be installed with the telescope's platform lift. With the telescope adapter installed, the total weight of the instrument is 675 pounds. This is about a factor of two below the nominal capacity of the rotator-guider module.

### 3.6 Detector Interface

The KPNO and CTIO Dewars are different designs, and as a result the KOSMOS and COSMOS exit plates are different. This difference is the only distinguishing feature between the two instruments. The KOSMOS Dewar and controller electronics are shown in Figure 3.

The Dewar does not mount directly to the exit plate. Instead, there is a mounting ring between the exit plate and Dewar that should make it more straightforward to install other Dewars in the future, and also served as an additional degree of freedom in case it was necessary to shim the Dewar to account for any optical misalignment.

The mounting ring includes a provision for rotational adjustment that allows the detector to be precisely aligned with the spectrograph slit, and to preserve that alignment when the Dewar is removed and reinstalled. There are also locating features that aid in the precise re-installation of the detector system at the same orientation.

## 4. ELECTRONICS AND CONTROL SYSTEM

The instrument electronics, control system, and instrument software are nearly identical to OSMOS. The only substantial difference is the shutter. KOSMOS and COSMOS use the model PS-500 by Sci-in Tech. This is a 5-inch square photometric aperture shutter with a pair of carbon fiber blades that provide even illumination across the field of view. The shutter allows exposure times as short as 0.01 seconds, although exposure times of 0.5 seconds or longer are recommended for accurate photometry.

## 5. DETECTOR SYSTEMS

Both instruments currently have an e2v 2048x4096 CCD (model 44-82) with a multi-layer anti-reflection coating optimized for broadband response. The KOSMOS detector is housed in a standard KPNO CCD Dewar, while the COSMOS CCD is housed in a custom Dewar whose design is adapted from the SOAR Optical Imager[6]. Both Dewars use liquid nitrogen to cool the detector.

The detector controller is an implementation of the basic NOAO "Monsoon" controller architecture[7,8,9] designated as "Torrent[10]". The Torrent controller design is a compact, low-power-consumption implementation capable of controlling a single CCD with multiple outputs. The Torrent controller also has design features intended to simplify maintenance support by observatory staff.

The read noise of the e2v detectors is approximately five electrons and the gain is 0.6 electrons per data number. Readout of the full detector takes 46 seconds. Several region of interest (ROI) modes are available that provide faster readout of a subset of the full array. Use of the ROI modes is a significant time saver for imaging and for target acquisition. Several binning options are also available.

The instrument design allows for the exchange of the CCD Dewar by the Observatory day crew. A red-sensitive LBNL CCD[11] has been purchased for both instruments and the Dewars and detector systems are under development.

# 6. USER INTERFACE

The user interface is an adaptation of the NOAO Observation Control System (NOCS)[12]. The software was originally developed for the NOAO wide-field IR Imager, NEWFIRM[13,14]. It has been adapted to the Mosaic wide-field optical imager[15], in addition to KOSMOS and COSMOS.

The NOCS operates by allowing the user to create scripts that are executed to perform a series of actions. For KOSMOS, scripts are generally fairly simple. For example:

- The instrument and telescope can be configured to obtain a specified number of calibration images (e.g. flat fields, arc spectra) with a particular instrument configuration.
- The instrument and telescope can be configured to obtain a specified number of science images (direct images or spectra) with a particular instrument configuration
- The images needed to align an object or field onto a slit or mult-slit mask can be obtained with a single script.

Scripts can be concatenated when the sequence of observations is certain to proceed without the need for interruption (for example, afternoon calibrations). The scripts can also be prepared in advance, which leads to more efficient operation. IRAF[16,17,18,19] and Python scripts are provided for efficient target acquisition, and IRAF is also available for inspection of the data as they are obtained.

The NOCS also supports several status displays. The most important displays for KOSMOS and COSMOS are:

- A status display for the Torrent controller and software
- A combined status display for the rest of the instrument and key telescope parameters
- A status display for the interface to the NOAO Data Handling System[20] (DHS).

These displays allow the observer to verify that the instrument is configured as desired, and to monitor the progress of observations.

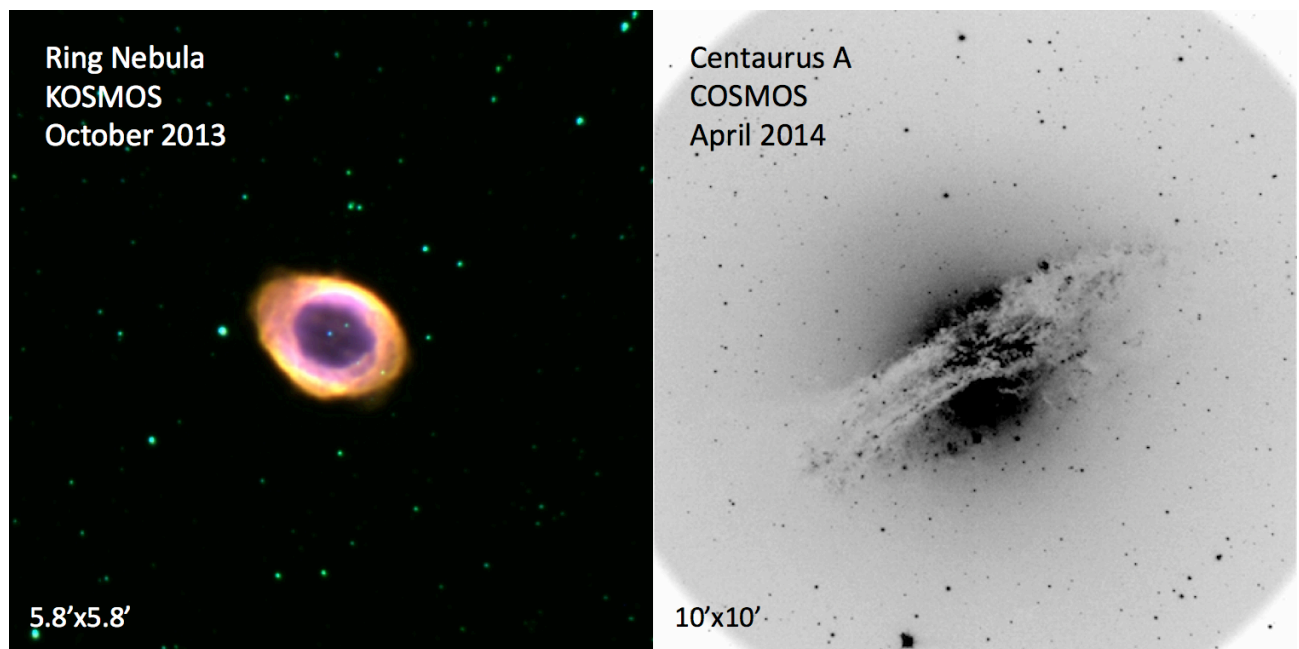

Figure 6. Images obtained with KOSMOS (left) and COSMOS (right) during their first commissioning runs. The left image is a multi-band color composite of the Ring Nebula (M57), while the right image is Centaurus A (NGC 5128) through an Hα filter.

# 7. COMMISSIONING AND PERFORMANCE

KOSMOS was commissioned in October 2013 and COSMOS was commissioned in April 2014. Basic commissioning of the instruments, in the sense of verifying functionality and interfaces, took approximately one clear night. Most of the commissioning time has been devoted to the development of efficient operation protocols.

## 7.1 Image Quality

Figure 6 shows images obtained with KOSMOS in October 2013 and COSMOS in April 2014. While the image quality was variable on these runs, images with FWHM better than one arcsecond have been recorded on sky.

The best measurements of the instrument image quality are from pinhole mask sequences that were obtained to focus the instrument with the instrument mounted on the telescope. The pinhole mask is mounted in the slit wheel and consists of a grid of 5μm diameter precision pinholes that cover the telescope focal surface. At the best focus, the FWHM of the pinholes are less than two pixels across the entire field of view. The pinhole mask was also used to verify the performance of each camera's fluid-coupled lens assemblies[4] in the lab prior to on-sky commissioning

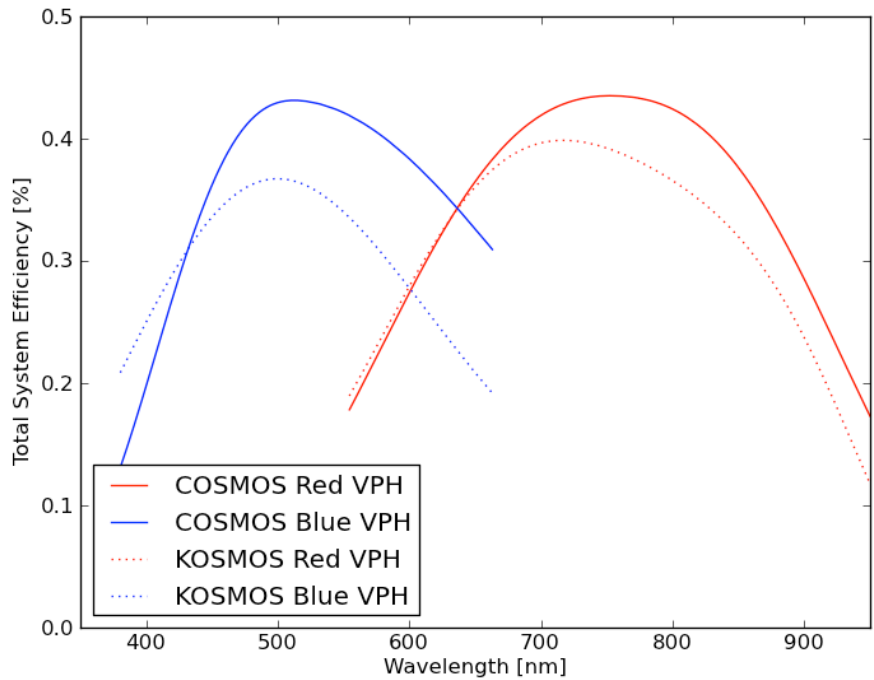

Figure 7. Measured total system efficiency of COSMOS (solid lines) and KOSMOS (dashed lines) with the Blue and Red VPH grisms. The efficiency was calculated from slitless observations of spectrophotometric standards near the zenith. The efficiency includes the telescope, optics, and the e2v CCD.

## 7.2 Throughput

Instrument throughput was measured with observations of standard stars near the zenith under photometric conditions. Spectra were measured without a slit, so the instrumental throughput did not need to be corrected for slit losses. Data for both instruments are shown in Figure 7. The throughput measurements include the instrument and the telescope.

The somewhat higher throughput for COSMOS is probably real, and if so, is likely due in large part to the different age of the primary mirror coatings. The Blanco primary was last coated in 2012, while the Mayall primary was last coated in 2009 (coating is planned for 2014).

## 7.3 Flexure

Flexure within the instrument can reduce efficiency in a variety of ways, and should ideally be minimal. The principal concerns related to flexure are degradation of resolution, if flexure during an exposure is significant, and degradation of flat-field corrections and loss of precision on wavelength zero-points, if flexure over the sky is significant. Measurements of KOSMOS are shown in Figure 8 for two different orientations of the instrument on the Mayall Cassegrain rotator. Measurements of COSMOS are very similar. The flexure is very well behaved and quite small, especially in the direction perpendicular to the slit. The flexure is approximately half the flexure measured for OSMOS[2], which was one of the design goals that motivated a stiffer optical bench design and focus actuators. The minimal flexure means that loss of resolution during the longest realistic exposures is not a concern, and flat field corrections and wavelength scales can be transferred from afternoon calibrations to night-time science observations. The flexure is sufficiently small that a wavelength zero-point correction, such as can readily be determined from bright sky lines, should suffice for most programs.

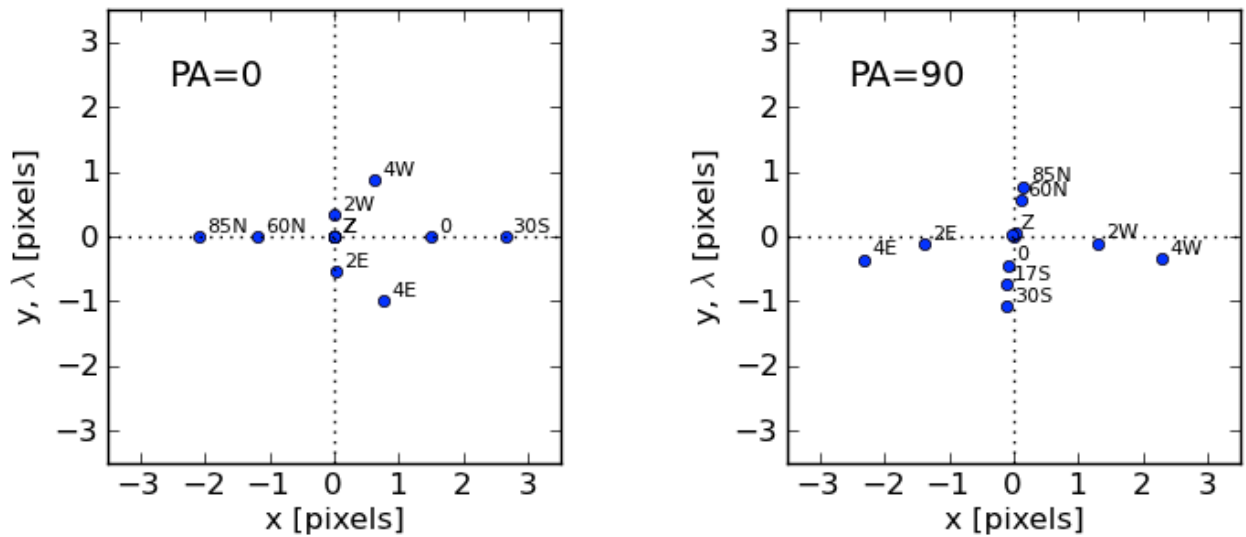

Figure 8. Measured KOSMOS image motion from December 2013 at PA=0 and PA=90 degrees. The points correspond to the mean pinhole centroid at various offset positions relative to the mean centroid positions at the zenith. The positions are offset 2 and 4 hours due east of zenith, 2 and 4 hours due west, +60 and +85 degrees north, on the celestial equator, and 30 degrees south. One pixel corresponds to 15μm or 0.29 arcseconds. The performance of COSMOS was measured in April 2014 with essentially identical results.

## 7.4 Spectroscopic Acquisition

The instruments do not have slit-viewing optics, and therefore the imaging mode is used to identify the target(s) and calculate the offsets required to place them on the slit(s). A fair amount of commissioning time was devoted to optimizing the efficiency of the acquisition process.

The basic steps are:

- Acquire a guide star in the offset guide probe field, and start guiding.
- Take an image of the target field through the slit or slit mask.
- Take an image of the target field without the slit mask. The same filter should be used for both fields; it does not have to be the filter that will be used for taking spectra. (The order of the two images does not matter, given the very good mechanical repeatability of the mechanisms.)
- Perform basic processing on the images (bias removal, combine sub-images from different amplifiers)
- Start a script to determine the offsets. This guides the user through the following steps:

    o Identify either the desired location on the long slit or alignment boxes in the multi-slit mask
    o Identify the target (for long-slit observations) or the alignment stars (for multi-slit observations)

- o The script calculates offsets in right ascension and declination. For multi-slit observations the script also calculates a rotation offset.
- If the offsets are more than a few arcseconds, a repetition of the acquisition sequence may lead to a small adjustment of the position. Smaller offsets are typically accurate enough that no further iteration is necessary.

The overall efficiency of this process is not determined by any one time-consuming step. Instead, it is necessary to work to optimize each step, and thereby save a fraction of a minute. For example, the acquisition images need only cover the region around the long slit (or the slit mask). Reading out only the region of interest reduces the detector readout and image processing times. The observer could prepare scripts or input subsequent steps while the preceding step is underway.

## 8. SUMMARY AND FUTURE

The KOSMOS and COSMOS instruments have been successfully commissioned at the Mayall and Blanco telescopes and are now available for use by the community. These instruments provide high efficiency spectroscopy with either a 10 arcminute long slit, or multiple slits in a slit mask arranged over 100 square arcminutes. KOSMOS and COSMOS may also be used as imagers over this same field of view with the facility 4-inch filters.

The instruments may have up to five dispersers mounted at once, and it is relatively straightforward to exchange dispersers during the night. While only two dispersers presently exist, several additional dispersers are planned for the near future.

## REFERENCES

[1] www.noao.edu/system/restar/
[2] Martini, P., Stoll, R., Derwent, M.A., et al., "The Ohio State Multi-Object Spectrograph," PASP, 123, 187-197 (2011).
[3] Stoll, R., Martini, P., Derwent, M.A., et al., "Mechanisms and Instrument Electronics for the Ohio State Multi-Object Spectrograph," Proc. SPIE 7735, 154-163 (2014).
[4] O'Brien, T.P., Derwent, M., Martini, P., Poczulp, G. "Design of the KOSMOS oil-coupled spectrograph camera lenses," Proc. SPIE 9151, *in press* (2014).
[5] Pogge, R.W., Atwood, B., Brewer, D.F., et al., "The multi-object double spectrographs for the Large Binocular Telescope", ed. I. S. McLean, S. K. Ramsay, H. Takami, Proc. SPIE, 7735, (2010).
[6] Walker, A. R., et al., *Instrument Design and Performance for Optical/Infrared Ground-based Telescopes,* ed. M. Iye, A. F. M. Moorwood, Proc. SPIE, 4841, 286-294 (2003).
[7] Starr, B. M., et al., *Scientific Detectors for Astronomy*, ed. P. Amico, J. W. Beletic and J. E. Beletic, Astrophysics and Space Science Lib., Vol. 300, Kluwer Academic Publishers, pp. 269-276 (2004).
[8] Sawyer, D., Moore, P. Rahmer, G. and Buchholz, N., in *Optical and Infrared Detectors for Astronomy,* ed. J. D. Garnett and J. W. Beletic, Proc. SPIE 5499, 489-498 (2004).
[9] www.noao.edu/nstc/monsoon/
[10] Hunten, M., Buchholz, N., George, R., Moore, P., Sawyer, D., *High Energy, Optical, and Infrared Detectors for Astronomy IV,* ed. A. D. Holland and D. A. Dorn, Proc. SPIE, 7742, 77421X (2010).
[11] Holland, S. E., et al., *High Energy, Optical, and Infrared Detectors for Astronomy II,* ed. D. A. Dorn and A. D. Holland, Proc. SPIE, 6276, 62760B (2006).
[12] Daly, P. N., *Software and Cyberinfrastruture for Astronomy,* ed. N. M. Radziwill and A. M. Bridger, Proc. SPIE 7740, 77403Q (2010).
[13] Autry, R. G., et al., *Instrument Design and Performance for Optical/Infrared Ground-based Telescopes,* ed. M. Iye, A. F. M. Moorwood, Proc. SPIE, 4841, 525-539 (2003).
[14] Daly, P. N., Buchholz, N. C., Fitzpatrick, M. J., Mills, D. Valdes, F. G., Zarate, N., Swaters, R., Probst, R. G., Dickinson, M., *Advanced Software and Control for Astronomy II.* ed. A. Bridger and N. M. Radziwill, Proc. SPIE 7019, 701913 (2008).
[15] Sawyer, D. G., Daly, P. N., Howell, S. B., Hunten, M. R., Schweiker, H., *Ground-based and Airborne Instrumentation for Astronomy III*, ed. I. S. McLean, S. K. Ramsay and H. Takami, Proc. SPIE, 7735, 77353 (2010).
[16] Tody, D., *Instrumentation in Astronomy VI,* Proc. SPIE 627, 733 (1986).

[17] Tody, D., *Astronomical Data Analysis Software and Systems II,* ed. R. J. Hanisch, J. V. Brissenden and J. Barnes, ASP Conf. Series, 52, 173 (1993).
[18] Fitzpatrick, M., *Astronomical Data Analysis Software and Systems XXI,* ed. P. Ballester, D. Egret and N. P. F. Lorente, ASP Conf. Series, 461, 595 (2012).
[19] www.iraf.net
[20] Fitzpatrick, M. J., *DTS: The NOAO Data Transport System*, Proc. SPIE 7737, 77371T (2010).